\documentclass[prl,twocolumn]{revtex4-1}

\usepackage{epsfig}
\usepackage{graphicx}
\usepackage{color}

\usepackage[caption=false]{subfig}

\begin{document}
\renewcommand{\thefigure}{\arabic{figure}}
\setcounter{figure}{0}

 \def\I{{\rm i}}
 \def\E{{\rm e}}
 \def\D{{\rm d}}

\bibliographystyle{apsrev}

\title{Rapid and reliable sky localization of gravitational wave sources}

\author{Neil Cornish}
\affiliation{Department of Physics, Montana State University, Bozeman, Montana 59717, USA}

\begin{abstract}
The first detection of gravitational waves by LIGO from the merger of two compact objects  has sparked new interest in detecting electromagnetic counterparts to
these violent events. For mergers involving neutron stars, it is thought that prompt high-energy emission in gamma rays and x-rays will be followed days to
weeks later by an afterglow in visible light, infrared and radio. Rapid sky localization using the data from a network of gravitational wave detectors is essential to
maximize the chances of making a joint detection. Here I describe a new technique that is able to produce accurate, fully Bayesian sky maps in seconds or less.
The technique can be applied to spin-precessing compact binaries, and can take into account detector calibration and spectral estimation uncertainties.
\end{abstract}

\maketitle

The detection by LIGO of the gravitational wave signal GW150914~\cite{Abbott:2016blz} heralds the beginning of the a new branch of astronomy. The information encoded in a gravitational wave signal
is highly complementary to that available from electromagnetic observations. There is considerable interest in detecting electromagnetic counterparts to
gravitational wave signals, as evidenced by the extensive follow-up campaign for GW150914~\cite{Abbott:2016gcq}, and the excitement generated by the tentative detection
of a gamma ray counterpart~\cite{Connaughton:2016umz}. High-energy counterparts to the sources detected by terrestrial interferometers are expected to be generated within seconds
of the peak gravitational wave emission, motivating the development of low-latency search and sky-mapping techniques~ \cite{Singer:2014qca, Berry:2014jja}. Notable recent developments
include reduced order models for the rapid generation of waveforms~  \cite{Cannon:2010qh, Field:2011mf, Canizares:2014fya, Blackman:2014maa}, a streamed
search for binary mergers~\cite{Cannon:2011vi}, and the {\em BayesStar} algorithm~\cite{Singer:2015ema} for producing low-latency sky maps (see also Ref.~\cite{Chen:2015eca}
for a similar approach).

Here I describe an efficient computational technique for producing low-latency sky maps based on a very fast and highly accurate approximation to the likelihood.
The new likelihood function is extremely cheap to evaluate, allowing for fully Bayesian sampling techniques to be employed to generate reliable sky maps
in seconds to minutes. By reliable I mean that $p$ percent of the time the signal will be found within the $p^{\rm th}$ credible interval of the sky map. Reliability
is critically important when allocating limited resources in the electromagnetic follow up campaign. 

The fast likelihood approach described here extends the stationary phase approximation (SPA) based approach for rapid evaluation of the likelihood~\cite{Cornish:2010kf, Cornish:2011ys}.
The method shares elements of the {\em BayesStar} algorithm~\cite{Singer:2015ema}, and the recently developed fast parameter estimation algorithm {\em FastPE}~\cite{Pankow:2015cra},
but has the advantage over {\em BayesStar} that it can handle systems where the orbital plane precesses due to spin-orbit and spin-spin coupling. The {\em FastPE} algorithm employs a harmonic
waveform expansion that can be applied to precessing systems, though only non-precessing systems were considered in the original study~\cite{Pankow:2015cra}.
One advantage of the approach described here is that it can easily incorporate calibration
uncertainties~ \cite{Abbott:2016jsd, Vitale:2011wu, farr} and uncertainties in the noise spectrum~  \cite{Littenberg:2013gja, Littenberg:2014oda}, which generally have a larger effect on
sky maps than marginalization over the intrinsic parameters (masses and spins) of the source. 

The production of rapid sky maps using the fast extrinsic likelihood technique would follow a low latency detection by the online compact binary searches, possibly followed up
with a fast Bayesian or maximum likelihood refinement of the intrinsic source parameters. The intrinsic waveform would then be used to set up the fast extrinsic likelihood calculation,
which can then be used to produce reliable sky maps via MCMC sampling.

The fast extrinsic likelihood calculation is best understood by starting with a simple example. Consider the dominant gravitational wave harmonic for a non-precessing quasi-circular inspiral.
Working in the frequency domain, the two polarization states are related such that $h_\times(f) = \I \epsilon h_+(f)$. The detector response is then
\begin{equation}
h(f) = h_+(f) (F_+  + \I \epsilon F_\times) \E^{2\pi \I f t_a}
\end{equation}
where $F_+(\alpha,\beta,\psi)$ and $F_\times(\alpha,\beta,\psi)$ are the antenna patterns for a detector, and $t_a$ is the arrival time relative to the geocenter.
Here $(\alpha,\beta)$ are the RA and DEC and $\psi$ is the polarization angle. The ellipticity $\epsilon$ is related to the inclination of the orbital angular momentum vector
to the line of sight, $\iota$, by
\begin{equation}
\epsilon = \frac{2 \cos\iota}{(1+\cos^2 \iota)} \, .
\end{equation}
We can re-express the antenna response as $F \E^{\I \lambda} = F_+  + \I \epsilon F_\times $
where
\begin{equation}
F = (F_+ ^2 + \epsilon^2 F_\times^2)^{1/2}
\end{equation}
and
\begin{equation}
\lambda = {\rm atan}(\epsilon F_\times/F_+)
\end{equation}
so that
\begin{equation}\label{sig}
h(f) = h_+(f) F \E^{\I \lambda} \E^{2\pi \I f t_a} \, .
\end{equation}
Adopting the standard Gaussian likelihood $p(d\vert h) \sim \E^{-\chi^2/2}$ we have
\begin{equation}\label{chi}
\chi^2 = (d - h \vert d - h) = (d \vert d) + F^2 ( h_+\vert h_+) - 2 (d\vert h) \, ,
\end{equation}
where $(a\vert b)$ denotes the noise spectrum $S_n(f)$ weighted inner product
\begin{equation}
(a\vert b) = \int \frac{ a^* b + b^* a}{S_n(f)} \, \D f \, .
\end{equation}
The inner products $D^2 = (d \vert d)$ and $H^2= ( h_+\vert h_+)$ do not depend on the extrinsic parameters $\alpha,\beta,\psi,\iota,t_a$ and have only to be computed
once for a given set of intrinsic parameters. These inner products also depend on the current estimates for the noise spectrum $S_n(f)$ and detector calibration.
The final term in the likelihood requires a little more work:
\begin{equation}
 (d\vert h)  =  F \E^{-\I \lambda} C(t_a) +   F \E^{\I \lambda} C^*(t_a)\, .
\end{equation}
where
\begin{equation}\label{Ce}
C(t_a) =   \int \frac{ h_+^*  d }{ S_n(f)} \,  \E^{2\pi \I f t_a} \D f  \, .
\end{equation}
We recognize that $C(t_a)$ has the form of an inverse Fourier transform, which can be evaluated rapidly using a Fast Fourier transform (FFT).  In order to have sufficient time
resolution (typically a tenth of a millisecond or less), it is necessary to zero-pad the frequency series prior to performing the FFT. Nonetheless, the computational cost is small.
Putting everything together we have
\begin{equation}\label{flike}
\chi^2 = D^2 + F^2 H^2  - 4 F \vert C(t_a) \vert \cos(\lambda - c(t_a))\, ,
\end{equation}
where we have written $C(t_a) =   \vert C(t_a) \vert \E^{\I c(t_a)} $. The cost of computing $D^2, H^2$ and $C(t_a)$ is typically a factor of 10 to 100 times that of computing the
likelihood directly (not including the waveform generation, which we assume has already been done). However, once the inner products are calculated, evaluating the likelihood
for new values of the extrinsic parameters is many orders of magnitude faster than the direct approach -- typically $10^4$ to $10^5$ times faster. The FastPE approach uses a
similar approach as Eqs. (\ref{chi}), (\ref{Ce}) to express the likelihood in terms of inner products between the signal model and the data. This fast likelihood can be used in
a standard Bayesian sampling scheme~\cite{Aasi:2013jjl, Veitch:2014wba}, such as Markov Chain Monte Carlo (MCMC), to generate sky maps in low latency. Below I explain
how the fast likelihood approach can be extended to include precession, calibration and spectral uncertainties, and even small variations in the extrinsic parameters (masses, spins {\it etc}).

\begin{figure}[htp]
\includegraphics[clip=true,angle=0,width=0.48\textwidth]{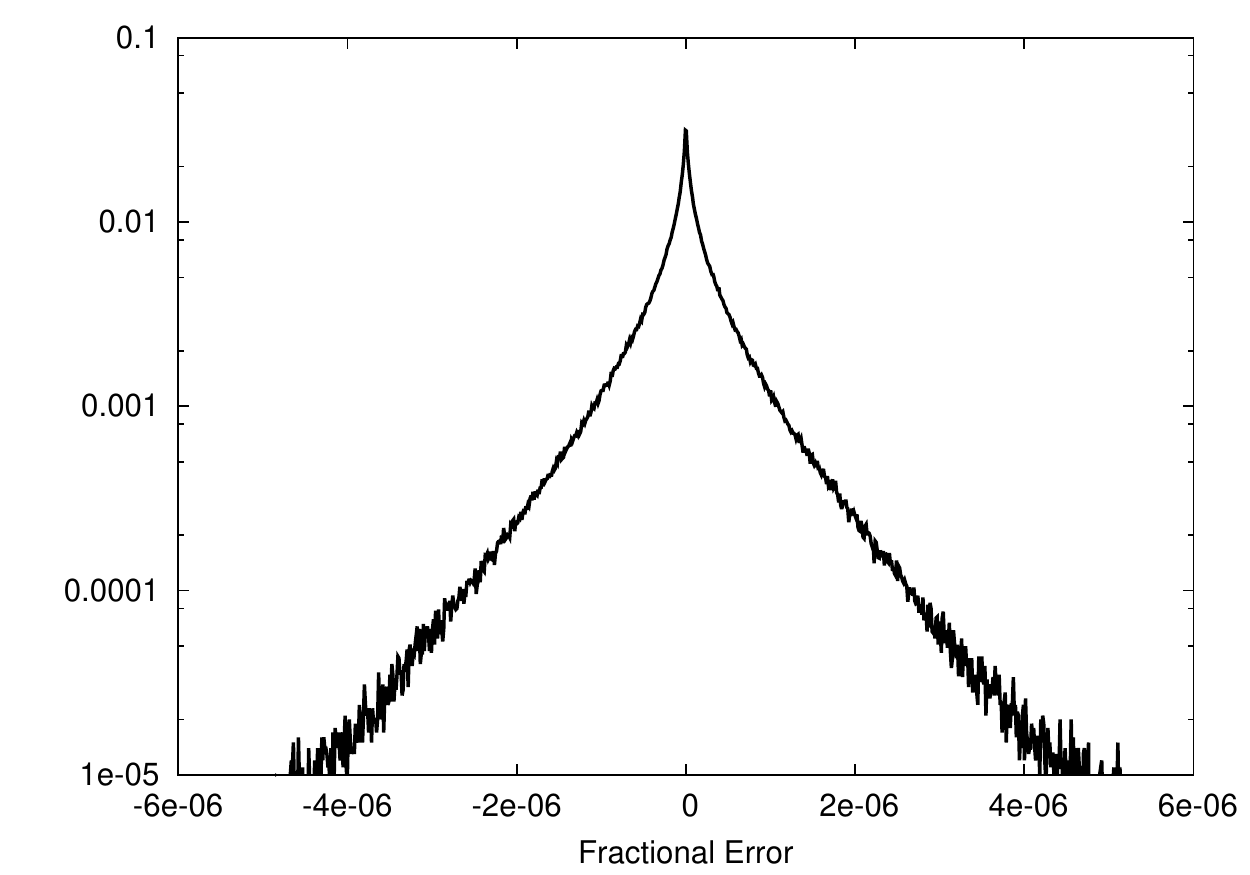} 
\caption{\label{fig:err} A histogram of the fractional error between the fast and standard likelihood calculation from a Monte Carlo sampling of extrinsic parameters
for a black hole merger signal.}
\end{figure}

Figure 1 shows a histogram of the fractional error in the log likelihood found by comparing the fast likelihood to the full likelihood calculation for a simulated black hole
merger. The histogram was produced by a Monte Carlo simulation that drew extrinsic parameters from the natural prior for the extrinsic parameters (uniform in location and orientation). Here the $C(t_a)$ function was generated
with samples spaced by $dt = 3\times 10^{-5}$ seconds, and linear interpolation was used between samples. Since the log likelihood scales as the amplitude signal-to-noise (SNR) ratio squared,
this level of accuracy is sufficient for signals with ${\rm SNR} < 100$. The time interval has to be reduced for louder signals.

\begin{figure}[htp]
\includegraphics[clip=true,angle=0,width=0.48\textwidth]{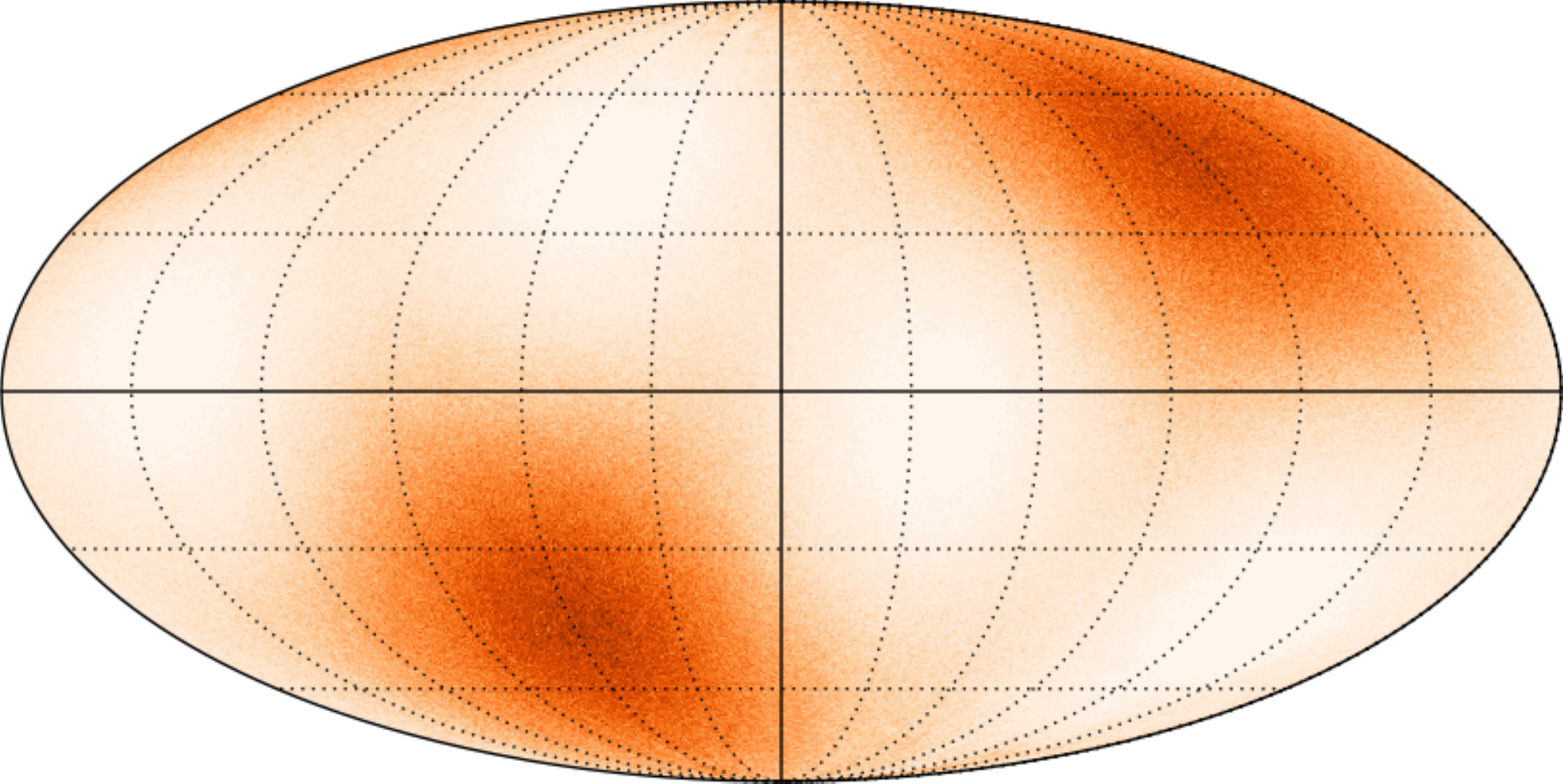} 
\includegraphics[clip=true,angle=0,width=0.48\textwidth]{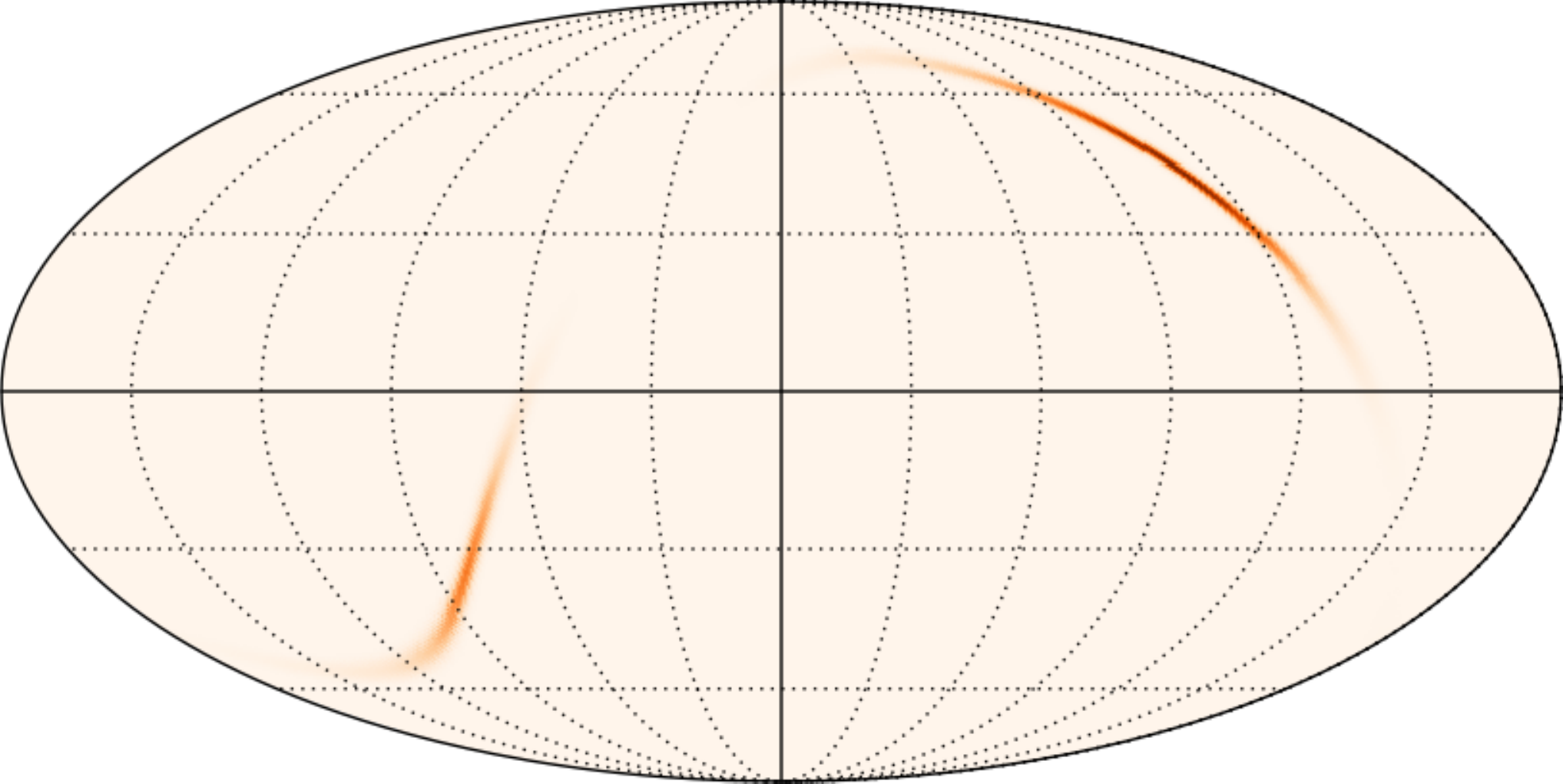} 
\includegraphics[clip=true,angle=0,width=0.48\textwidth]{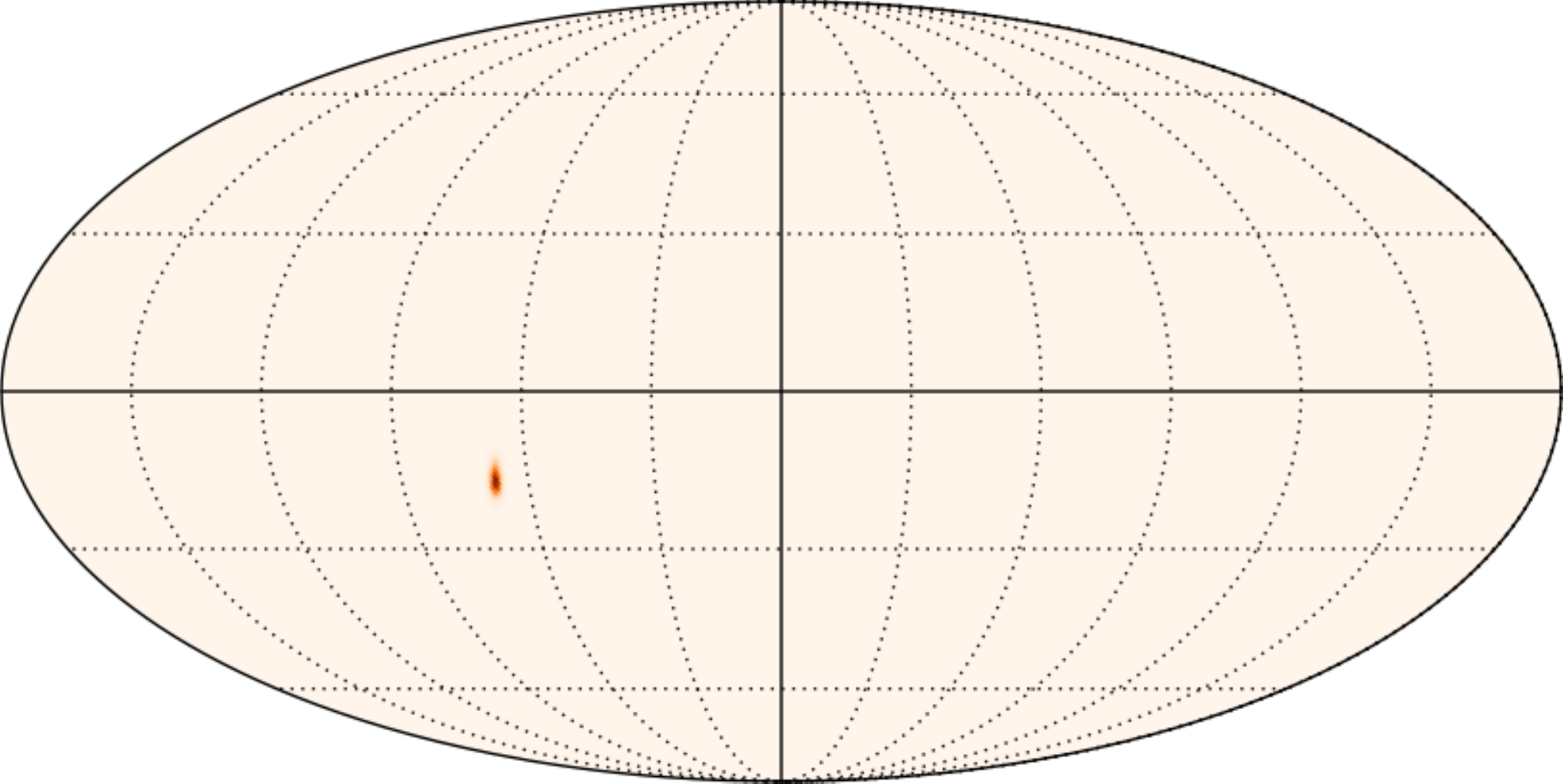} 
\caption{\label{fig:skies} Fast sky maps showing the posterior distribution for the sky location of a simulated binary black hole merger. The upper panel is with just the Hanford detector,
the middle panel is for the Hanford/Livingston network and the lower panel is for the Hanford/Livingston/Virgo network. Each map took just a few seconds to produce on a single computational core.}
\end{figure}

Figure 2 shows examples of sky maps produced using the fast likelihood function (\ref{flike}) and a standard MCMC sampling routine (taken from the {\em BayesWave} algorithm~\cite{Cornish:2014kda}).
A simulated signal for a spin-aligned black hole merger with masses $m_1 = 20\, M_\odot$ and  $m_2 = 15\, M_\odot$ was added to simulated Gaussian noise for the LIGO Hanford (H)
and Livingston (L) detectors and the Virgo (V) detector, each using the zero-detuning, high-power aLIGO design noise spectral density~\cite{AdvLIGO-noise}
The signal-to-noise ratios in each detector were ${\rm SNR}_{\rm H} = 9.9$,  ${\rm SNR}_{\rm L} = 8.9$ and
${\rm SNR}_{\rm V} = 4.5$.  Each map was produced from $10^7$ independent samples, and took a few seconds to produce. The advantage of having a large number of independent samples
is that regions of low probability are accurately mapped, so that even the 95\% credible interval is well defined.

The method described above can be applied to more complicated signals. For signals that include multiple harmonics: $h_+ = \sum_k h_{+k}$, $h_\times(f) = \I \sum_k \epsilon_k h_{+k}(f)$, the
derivation is unchanged so long as there is little overlap between the harmonics: $\vert (h_j \vert h_k) \vert /\sqrt{(h_j \vert h_j) (h_k \vert h_k)} \ll 1$ for $i\ne j$. Systems with mis-aligned
spins pose more of a challenge since the resulting precession of the orbital plane make $F$ and $\lambda$ functions of frequency. However, because the
precession timescale is long compared to the orbital time scale, they vary slowly in frequency. Here I am considering signals where the SPA can be used to map between time and frequency.
It is possible to expand $F^2$, $F\cos\lambda$ and $F\sin\lambda$ in a suitable basis,
such as Chebyshev polynomials of the first kind, $T_n(f)$. The number of terms needed can be reduced by expressing the expansion in terms of $\phi_z(f)$, the precession angle through which the orbital
angular momentum rotates around the total angular momentum vector. The series converges more rapidly with this parameterization since the oscillations are roughly evenly spaced. An expansion in terms of $\log f$ also converges quickly.
Writing $F^2= \sum_k \alpha_k T_k(\phi_z(f))$, the $(h\vert h)$ term becomes
\begin{equation}\label{hh}
H^2 = (h \vert h)  = \sum_k \alpha_k  H_k \,
\end{equation}
where the constants $H_k$ are given by
\begin{equation}
H_k =  \int \frac{  2 h^* h \, T_k(\phi_z(f)) }{ S_n(f)} \, \D f  \, .
\end{equation}
These constants have only to be evaluated once. The $(d\vert h)$ integral now requires the evaluation of
\begin{equation}
F_c(t_a) =\Re \left\{  \int \frac{ F \cos \lambda \, h_+^* d }{ S_n(f)} \, \E^{2\pi \I f t_a} \, \D f \right\} \, ,
\end{equation}
and
\begin{equation}
F_s(t_a) =  \Im \left\{  \int \frac{ F \sin \lambda \, h_+^* d }{ S_n(f)} \, \E^{2\pi \I f t_a} \, \D f \right\} \, .
\end{equation}
Expanding $F \cos\lambda= \sum_k \beta_k T_k(\phi_z(f))$ and $F \sin\lambda= \sum_k \gamma_k T_k(\phi_z(f))$ we have
\begin{equation}
F_c(t_a) =  \sum_k \beta_k \Re C_k(t_a), \quad F_s(t_a) =  \sum_k \gamma_k \Im C_k(t_a),
\end{equation}
where each
\begin{equation}
C_k(t_a) =  \int \frac{ T_k(\phi_z(f)) h_+^*  d }{ S_n(f)} \, \E^{2\pi \I f t_a}\, \D f  \, 
\end{equation}
is computed by an inverse FFT. The log likelihood for time delay $t_a$ is then given by $-\chi^2/2$ where
\begin{equation}\label{cheblike}
\chi^2 =  D + \sum_k \left( \alpha_k H_k -2( \beta_k   \Re C_k(t_a)  +\gamma_k \Im C_k(t_a)) \right) \, .
\end{equation}

Precession drives up the computational cost by factors of tens or hundreds, depending on how many precession cycles occur in the sensitive band of the
detectors, and on how large the precession effects are. Figure 3 shows a fairly extreme example for a highly precessing black hole binary with masses $m_1 = 10\, M_\odot$ and  $m_2 = 5\, M_\odot$ and
mis-aligned spins with dimensionless magnitudes $\chi_1 = 0.7$ and $\chi_2 = 0.5$. Using $N=100$ terms in the Chebyshev expansion recovers the precession dynamics to better than one percent accuracy.
The convergence is exponential in $N$. The Chebyshev expansion coefficients and the reconstructed polynomials can be computed at ${\cal O}(N \log N)$ cost using a FFT.

\begin{figure}[htp]
\includegraphics[clip=true,angle=0,width=0.48\textwidth]{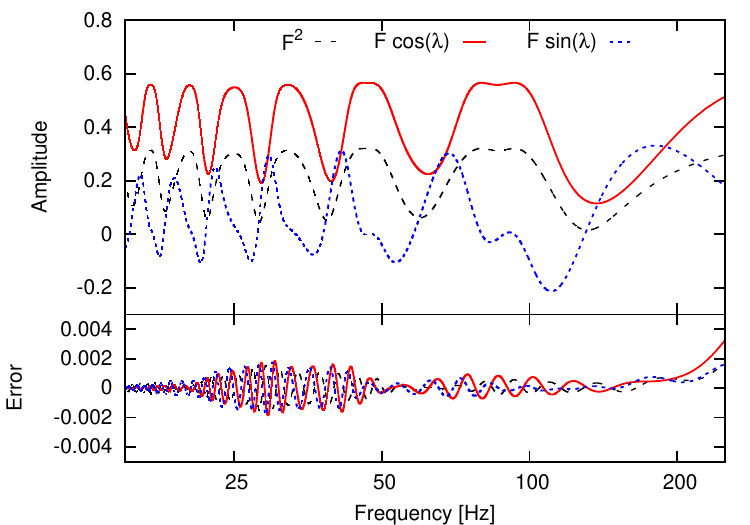} 
\caption{\label{fig:pre} The upper panel shows the impact of precession on the three extrinsic factors that enter the likelihood. The lower panel shows the error in recovering the evolution
using a Chebyshev polynomial expansion with 100 terms.}
\end{figure}

Being able to include spin precession effects in the fast likelihood calculation will be particularly important for black hole/neutron star binaries as these systems are in-band for minutes, reaching high
frequencies at merger, and undergoing many precession cycles, making the direct calculation of the likelihood very costly. The fast likelihood calculation can be incorporated into the standard
Bayesian inference codes used by the LIGO/Virgo collaboration, allowing for large numbers of fast extrinsic parameter updates to be performed following each slow intrinsic parameter update.
The approach described above is particularly easy to apply to effective models that incorporate precession effects by rotating a non-precessing
waveform~\cite{Hannam:2013oca, Boyle:2011gg, O'Shaughnessy:2011fx, Arun:2008kb}.

Marginalization over calibration and spectral models can be incorporated in the fast likelihood calculation using the same approach as for precession. Calibration errors enter the observed signal
$h_{\rm obs}$ as frequency dependent amplitude and phase errors~\cite{Vitale:2011wu}:
\begin{equation}
h_{\rm obs} =  h(f) (1+ \delta A(f))  \E^{\I \delta \phi(f)} \, .
\end{equation}
These can be accommodated by replacing $F \rightarrow F (1+ \delta A(f))$ and $\lambda \rightarrow \lambda +  \delta \phi(f)$ in the derivation given above for precessing systems.
The only difference is that the $F$ and $\lambda$ functions are different in each detector, while the precession effects are common to all the detectors. Note that marginalization
over the calibration uncertainties adds almost no additional cost to the analysis of precessing systems. Similarly, marginalization over the spectral model~\cite{Littenberg:2014oda} can be expressed as
\begin{equation}
\frac{1}{S_n(f)} = \frac{B(f)}{S_{n, {\rm ref}}(f)}
\end{equation}
where $S_{n, {\rm ref}}(f)$ is a reference spectrum, and $B(f)$ is a frequency dependent scaling. Again, $B(f)$ can be absorbed into the $F$ function for the $(h\vert h)$ and $(d\vert h)$ terms, while
an addition expansion of $B(f)$ is needed to compute the $(d\vert d)$ term using the same method described in (\ref{hh}).

Going one step further, for SPA waveforms that can be written: $h(f,\vec{\theta}) = A(f,\vec{\theta}) \exp(\I \Phi(f,\vec{\theta}))$, where $\vec{\theta}$ denotes the intrinsic parameters of the source, the
fast likelihood approach can be extended to cover intrinsic parameters using the Chebyshev expansion. The key to making this possible is that waveforms that yield likelihoods that have any chance of being accepted in a MCMC exploration of parameter space have amplitudes and phases that are close to those of the reference waveform used to compute the $H_k$ and $C_k(t_a)$~\cite{Cornish:2010kf}.

\begin{figure}[htp]
\begin{center}$
\begin{array}{cc}
\includegraphics[clip=true,angle=0,width=0.24\textwidth]{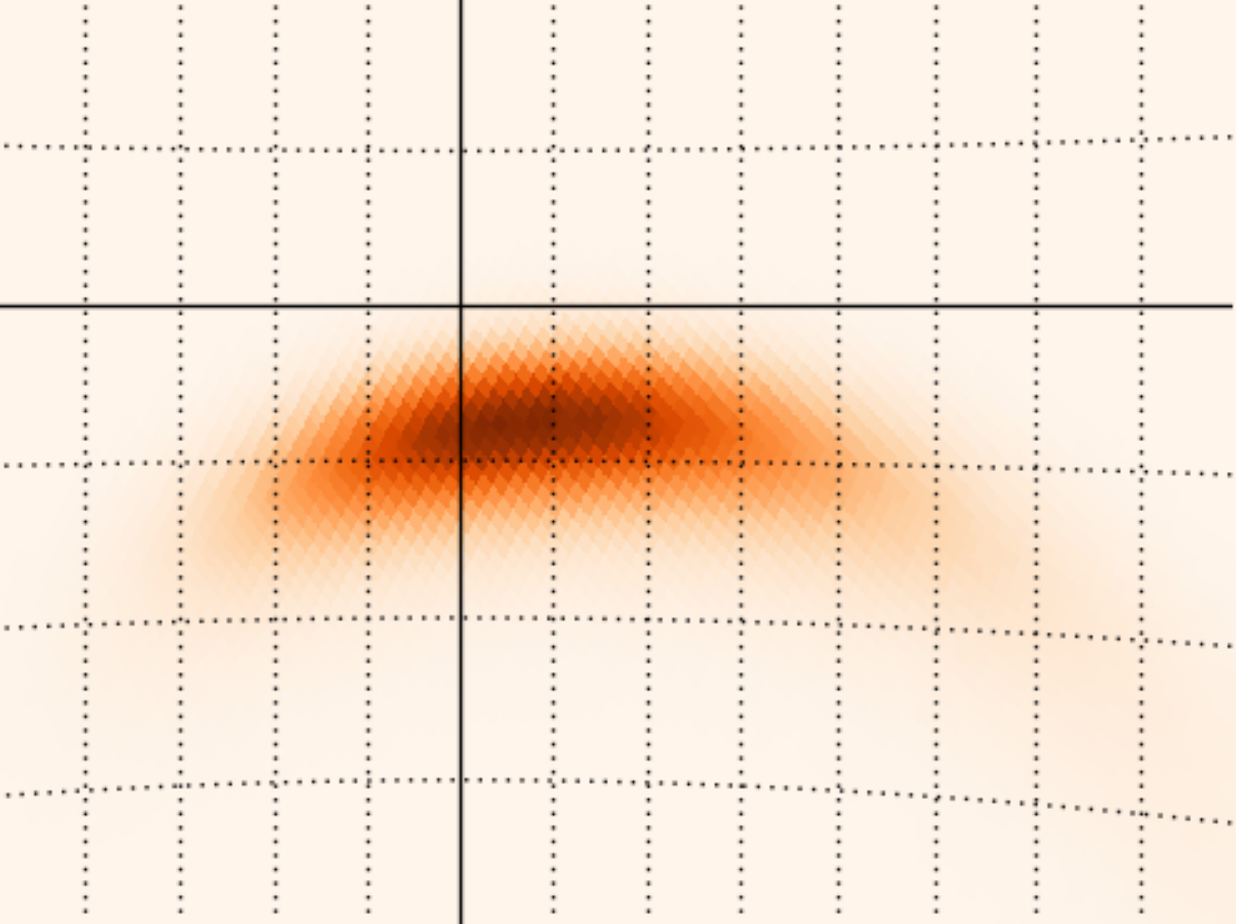} &
\includegraphics[clip=true,angle=0,width=0.24\textwidth]{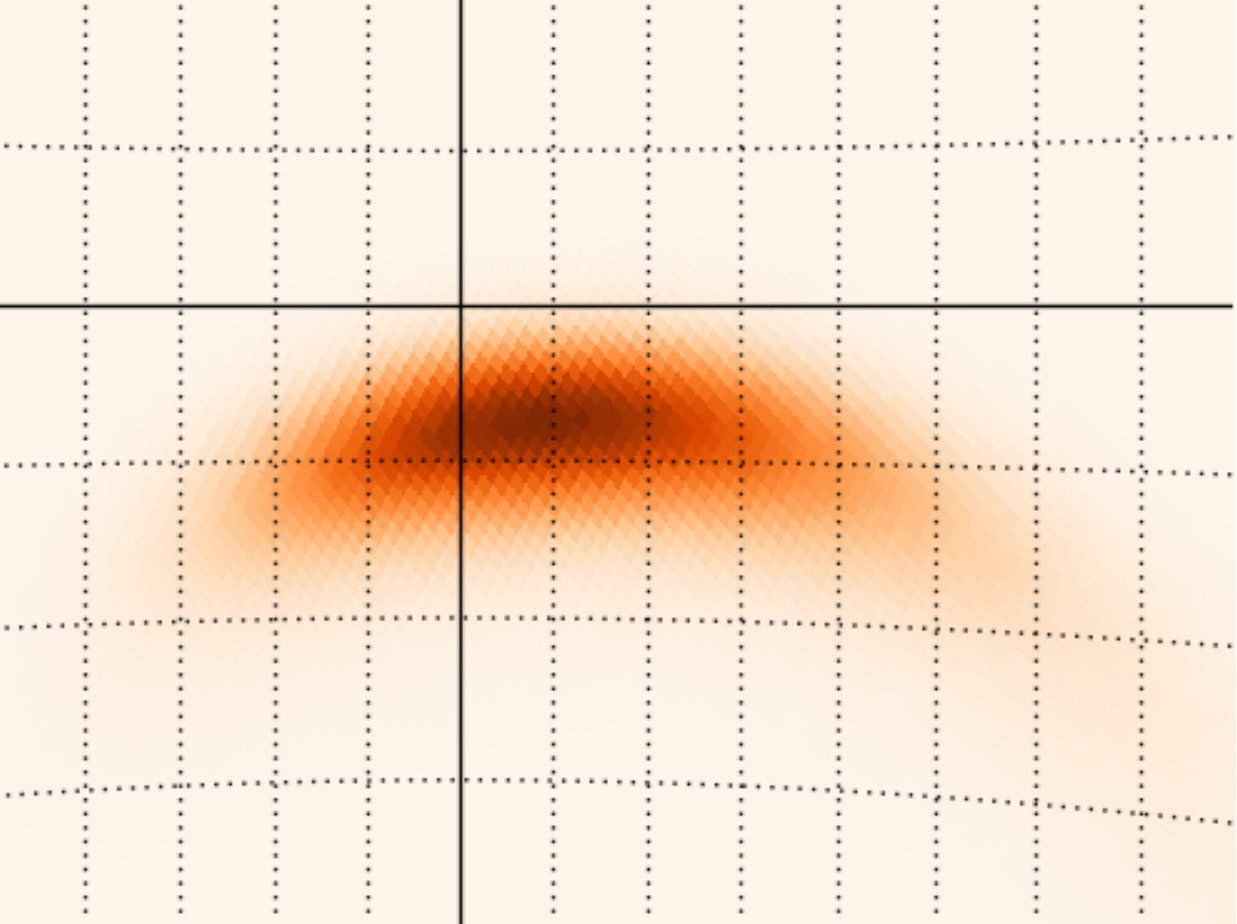} 
\end{array}$
\end{center}
\caption{\label{fig:cal} Gnomonic projections of sky maps for a simulated LIGO Hanford/Livingston observation of a binary system with the same intrinsic parameters as used in Figure 1, but
with a different sky location and ${\rm SNR}=19.7$. The left panel is for fixed calibration, while the right panel includes marginalization over calibration errors. The dark lines in each figure mark
$\rm{(RA,DEC)} = (140^\circ,40^\circ)$. The dotted meridians are space by $3^\circ$ and the parallels by $5^\circ$. }
\end{figure}

Figure 4 illustrates the application of the Chebyshev expanded likelihood (\ref{cheblike}) for producing sky maps that are marginalized over calibration errors. The amplitude and phase errors in each detector
were modeled as a Gaussian process in $\log f$ with zero mean and standard deviation $\sigma_A = 4\%$ and $\sigma_\phi = 4^\circ$, and a correlation scale chosen to give an average of 3 zero crossings
in the sensitive band of the detectors (similar to the model used in Ref.~\cite{Vitale:2011wu} and consistent with the uncertainties found during the first aLIGO run~\cite{Abbott:2016jsd}). The simulation is for
a two detector LIGO Hanford/Livingston network for a non-precessing source with the same intrinsic parameters are used in Figure 1, but with a different sky location and ${\rm SNR}=19.7$. The calibration
uncertainties shifted the maximum a posteriori sky location by $2.5^\circ$ and expanded the 90\% credible region from $254\; {\rm deg}^2$ to $291\; {\rm deg}^2$.

Summary: The current state-of-the art Bayesian parameter estimation techniques for analyzing LIGO/Virgo data~\cite{Veitch:2014wba} take days or weeks to produce accurate sky maps. Using the techniques
I have described, it is now possible to produce accurate sky maps in seconds or minutes, not just for simple systems under ideal conditions, but for fully precessing binary mergers with full marginalization
over calibration and spectral uncertainties. The fast extrinsic likelihood calculation has already been implemented in the {\em BayesWave}~\cite{Cornish:2014kda} transient analysis pipeline, and will be used during the second
advanced LIGO observing run starting in late 2016.

\section*{Acknowledgments}
This work was supported by NSF award PHY-1306702. Helpful feedback on an early draft was provided by Chris Pankow, Richard O'Shaughnessy and Albrecht R\"udiger.

\end{document}